\documentclass[11pt,twoside]{article}
\usepackage{macro-fsut-eng}
\usepackage{graphicx}

\usepackage[T1]{fontenc} 

\usepackage{latexsym}
\usepackage{verbatim}

\begin{document}

\vskip 1.0cm
\markboth{L. Amarilla and E.F. Eiroa}{Shadow cast by a Kaluza-Klein spinning dilaton black hole}
\pagestyle{myheadings}

\vspace*{0.5cm}
\title{Shadow cast by a Kaluza-Klein spinning dilaton black hole}

\author{Leonardo Amarilla,$^1$ and Ernesto F. Eiroa,$^{2,1}$}
\affil{$^1$Departamento de F\'isica, FCEN-UBA, Buenos Aires, Argentina.\\
$^2$Instituto de Astronom\'ia y F\'isica del Espacio, Buenos Aires, Argentina}

\begin{abstract}
We examine the shadow of a rotating Kaluza-Klein black hole in Einstein gravity coupled to a Maxwell field and a dilaton. The size and the shape of the shadow depend on the mass, the charge, and the angular momentum of the compact object. For a given mass, the size increases with the rotation parameter and decreases with the electric charge. The distortion with respect to the non rotating case grows with the charge and the rotation parameter. For fixed values of these parameters, the shadow is slightly larger and less deformed than in the Kerr-Newman case.
\end{abstract}

\section{Introduction}
\label{sec:introduction}

The study of gravitational lensing by black holes has received great attention in the last decade, due to the evidence of the presence of supermassive compact objects at the galactic centers. The apparent shapes (or shadows) of non-rotating black holes are circular, but rotating ones present a deformation produced by the spin; topic recently investigated by several researchers, both in Einstein gravity and in modified theories, with the expectation that direct observation of black holes will be possible in the near future. The analysis of the shadows will be a useful tool for obtaining properties of astrophysical black holes and comparing different gravitational theories. Here we consider the shadow cast by a Kerr and by a Kaluza-Klein rotating dilaton black hole with charge, corresponding to a coupling parameter $\gamma =\sqrt{3}$. We pay special attention to the analysis of the shadow of the supermassive black hole in the  center of our galaxy.

\section{The Kerr case}
\label{sec:kerr}

In Boyer-Lindquist coordinates, the Kerr solution has the line element (G=c=1)
\begin{eqnarray} \label{kerr}
ds^{2}&=&-( 1 - \frac{2Mr}{\Sigma}  )dt^{2} - \frac{ 4 M a r \sin^{2}\theta }{ \Sigma } d\varphi dt + \frac{\Sigma}{\Delta}dr^{2}+\Sigma d\theta^{2} \nonumber \\
&&+\left[ (r^{2}+a^{2})^{2}-\Delta a^{2}\sin^{2}\theta\right] \frac{\sin^{2}\theta }{ \Sigma } d\varphi^{2},
\end{eqnarray}
where $\Sigma=r^{2}+a^{2}\cos^{2}\theta$, $\Delta=r^{2}-2Mr+a^{2}$, and $a=J/M$ is the rotation parameter. The horizons are obtained by solving the equation $\Delta =0$, which gives $r_{\pm}=M^{2} \pm \sqrt{M^{2}-a^{2}}$, where $r_{+}$ is the outer (event) horizon and $r_{-}$ is a Cauchy inner horizon. Kerr spacetime is stationary and axisymmetric. These symmetries have associated Killing vectors, so $p_{t}=-E$ y $p_{\varphi}=L_{z}$ are conserved along the geodesic movement of a particle. There is an additional hidden symmetry associated to a fourth conserved quantity, the Carter constant (Chandrasekhar 1992). The geodesic structure is determined from the Hamilton-Jacobi equation:
\begin{equation}  \label{SHJ1}
\frac{\partial S}{\partial \lambda}=-\frac{1}{2}g^{\sigma\nu}\frac{\partial S}{\partial x^{\sigma}}\frac{\partial S}{\partial x^{\nu}},
\end{equation}
where $\lambda$ is an affine parameter along the geodesics, $g_{\sigma \nu}$ is the metric tensor, and $S$ is the Jacobi action. When the problem is \textit{separable}, $S$ can be written in the form
\begin{equation}  \label{SHJ2}
S=\frac{1}{2}\mu ^2 \lambda - E t + L_z \varphi + S_{r}(r)+S_{\theta}(\theta),
\end{equation}
where $\mu $ is the particle mass. The equations of motion result from $p_{\nu}=\partial S / \partial x^{\nu}$. In the case of photons ($\mu =0$), one can obtain that
$$
\Sigma\frac{dt}{d\lambda }=\frac{1}{\Delta}\left[(r^{2}+a^{2})^{2}-\Delta a^{2} \sin^{2}\theta-2Mar\xi\right], \;\;\;\;\;\;\;\;
\Sigma\frac{dr}{d\lambda }=\sqrt{\mathcal{R}}, 
$$
$$
\Sigma\frac{d\theta}{d\lambda }=\sqrt{\Theta}, \;\;\;\;\;\;\;\;
\Sigma\frac{d\varphi}{d\lambda }=\frac{1}{\Delta}\left[2Mar+(\Sigma-2Mr)\xi \csc^{2}\theta  \right].
$$
with
\begin{equation}\label{R}
\mathcal{R}(r)=\left[(r^2+a^2)E -a L_z \right]^2-\Delta\left[\mathcal{K}+(L_z -a E)^2\right],
\end{equation}
\begin{equation}\label{theta}
\Theta(\theta)=\mathcal{K}+\cos^2\theta\left[a^2E^2-L_{z}^{2}\csc^2\theta \right].
\end{equation}
The unstable orbits of photons with constant $r$ satisfy the conditions
$\mathcal{R}=0$ and $d\mathcal{R}/dr=0$. For a Kerr black hole, the radius solution to this system of equations depends on the trajectory of the photon. One can establish a relation between this radius and the conserved quantities, so the system of equations can be solved for the impact parameters $\xi=L_z/E$ y $\eta=\mathcal{K}/E^2$. The physical solution is given by
\begin{equation}\label{xietak}
\xi (r)=\frac{(r^{2}-a^{2})M - \Delta r}{a(r-M)},
\;\;\;\;\;\;\;\;
\eta (r) =\frac{r^{3}\left[4 M \Delta - r (r-M)^{2}\right]}{a^{2}(r-M)^{2}}.
\end{equation}
The parameters $\xi $ and $\eta $ above correspond to the unstable photon orbits. The apparent position of the photon sphere in the sky of  a distant observer, generates the contour of the shadow. The ``celestial'' coordinates of the observer  are defined as follows: $\alpha$ is the apparent perpendicular distance of the image as seen from the axis of symmetry, and $\beta$ is the apparent perpendicular distance of the image from its projection on the equatorial plane. For an angle of observation $\theta_{0}$ with respect to the equatorial plane, they take the form (V\'azquez et al. 2004)
\begin{equation}  \label{alphabeta}
\alpha=-\xi\csc\theta_{0}
\;\;\;\; \mathrm{and} \;\;\;\;
\beta=\pm \sqrt{\eta + a^{2}\cos ^{2}\theta_{0}-\xi^{2}\cot ^{2}
\theta_{0}}.
\end{equation}
In the Schwarzschild case ($a=0$) the apparent shape of the photon sphere is a circle of radius $3\sqrt{3}M$, while for $a\neq 0$ the contour has an asymmetric form because co-rotating photons interact with a more intense potential than counter-rotating ones, producing a closer approach to the black hole in the former case. An observer located in the equatorial plane of the object ($\theta_{0}=\pi/2$) is natural when considering the Galactic center supermassive black hole, in this case Eqs. (\ref{alphabeta}) take the form $\alpha=-\xi$ and $\beta=\pm \sqrt{\eta}$. Besides, the effects of rotation  are stronger when seen from this plane.  For more details, see for example Chandrasekhar book (1992).

\section{Kaluza-Klein rotating black hole}
\label{sec:kkrbh}

The action corresponding to standard gravity coupled to the Maxwell field $F^{\sigma \nu}$ and the (scalar) dilaton field $\phi$, in units such that $16\pi G=c=1$, reads
\begin{equation}
\mathcal{S} =\int d^4x\sqrt{-g}\left[ -R+2(\nabla \phi)^2+e^{-2\gamma \phi }F^2\right] ,
\label{emd}
\end{equation}
where $R$ is the Ricci scalar. Exact stationary rotating solutions are only known for certain values of the coupling parameter; $\gamma =\sqrt{3}$ leads to the so called Kaluza-Klein rotating black hole, which is obtained by taking the product of the four dimensional Kerr metric, in the Boyer-Lindquist coordinates, with an extra dimension possessing translational symmetry, and then making a boost transformation with velocity $v$ along the fifth dimension. The four dimensional section has the form  (Frolov et al. 1987; Horne et al. 1992)
\begin{eqnarray} \label{metric}
ds^2  &=& -\frac{1-Z}{B}dt^{2}-\frac{2 a Z \sin^{2}\theta}{B\sqrt{1-v^{2}}}dt d\varphi + \frac{B\Sigma}{\Delta_{0}}dr^{2} + B\Sigma d\theta^{2}  \nonumber \\ 
&& + \left[ B(r^{2}+a^{2})+a^{2}\frac{Z}{B}\sin^{2}\theta \right] \sin^{2}\theta d\varphi^{2},
\end{eqnarray}
where $\Sigma=r^2+a^2 \cos^2\theta $, $\Delta_{0} = r^2 -2 m r + a^2$, $B=\sqrt{1+Zv^{2}/(1-v^{2})}$,  and $Z=2 m r/\Sigma$, with $ m $ the mass and $a$ the rotation parameter of the original Kerr solution. The metric (\ref{metric}), along with the electromagnetic vector field  $A_{t}=(1/2)v(1-v^{2})^{-1}ZB^{-2}$, and $A_{\varphi}= - (1/2)av(1-v^{2})^{-1/2}ZB^{-2}\sin^{2}\theta$, and the dilaton field $\phi=-(\sqrt{3}/2)\log B$, is a solution of the equations of motion corresponding to the action (\ref{emd}) for $\gamma=\sqrt{3}$. The geometry (\ref{metric}) is asymptotically flat and represents a black hole with physical mass $M=m\left[ 1+ (1/2) v^{2}(1-v^{2})^{-1} \right] $, charge $Q=m v(1-v^{2})^{-1}$, and angular momentum $J=m a(1-v^{2})^{-1/2}$. The physical rotation parameter is defined by $A=J/M$. The sign of the charge $Q$ is determined by the sign of $v$, due to the physical bound $|v|<1$; if $v=0$ one recovers the Kerr solution. The roots of $\Sigma$ and $\Delta_{0}$ are associated to a curvature singularity at $r=0$ and $\theta=\pi/2$, and to regular horizons, respectively. The event horizon is located at $r_{+}= m + \sqrt{m^2 - a^2}$, and exists if $m^{2}\geq a^{2}$. We adopt $M=1$, which is equivalent to adimensionalize all quantities with $M$.

From the Hamilton-Jacobi equation, for null geodesics ($\mu  =0$) one can obtain the    corresponding equations of motion (Amarilla et al. 2013):
$$
B\Sigma \frac{dt}{d\lambda}=\frac{2 m r}{\Delta_{0}}\left[(r^{2}+a^{2})\left(\frac{1}{Z}+\frac{v^{2}}{1-v^{2}} \right) E + a^{2}E\sin^{2}\theta -\frac{a L_z}{\sqrt{1-v^{2}}} \right] ,
$$
$$
B\Sigma\frac{dr}{d\lambda}=\sqrt{\mathcal{R}},
\;\;\;\;\;
B\Sigma\frac{d\theta}{d\lambda}=\sqrt{\Theta},
\;\;\;\;\;
B\Sigma \frac{d\varphi}{d\lambda}=\frac{2 m r}{\Delta_{0}}\left(\frac{a E}{\sqrt{1-v^{2}}} -\frac{Z-1}{Z} L_z \csc^{2}\theta \right),
$$
where the function $\mathcal{R}(r)$ has the form
\begin{equation}
\mathcal{R}=\mathcal{R}_{\mathrm{Kerr}}+\frac{2r\left\{ \left[ (a E-L_z)^{2}-2 L_z^{2}-\mathcal{K}+2E^{2}r^{2}\right] v^{2}  +4 a L_z E (1-\sqrt{1-v^{2}})  \right\}}{2-v^{2}},
\end{equation}
and $\Theta(\theta)$ is given by Eq. (\ref{theta}). The photon sphere conditions $\mathcal{R}(r)=0$ and $d\mathcal{R}(r)/dr=0$ are fulfilled by the impact parameters 
\begin{eqnarray} \label{eqxieta}
\xi(r)&=&\frac{2(a^{2}-r^{2})\sqrt{1-v^{2}} + \Delta_{0} \sqrt{\vartheta } }{\zeta } ,
\nonumber \\
\eta(r)&=&\frac{r^2}{\zeta ^2}  \left\{ 4 \Delta_{0} \sqrt{\vartheta } \sqrt{1 - v^2} + 2 a^2 r (2-v^2)^2  + 4 a^2 v^2(1 -  v^2) \right. \nonumber \\
&&  - \frac{r}{2-v^2} \left[8 r (5 + (r-4) r) + 4 (8 - r (31 +3(r-6)r)) v^2 \right. \nonumber \\
&& \left. \left. + 2 (-32 + r (58 + 3 (r-8) r)) v^4 - (r-4)^2 (r-2) v^6 \right]   \right\},
\end{eqnarray}
where $\vartheta= r[2v^{2}+r(2-v^{2})](2-v^{2})$ and $\zeta = a [2(1-v^2) - r(2- v^2)]$. The  celestial coordinates of the contour of the shadow are determined by replacing these equations in Eqs. (\ref{alphabeta}), which are also valid for this spacetime. For an equatorial observer ($\theta _{0}=\pi/2$) these equations simplify to $\alpha=-\xi$ and $\beta=\pm \sqrt{\eta}$. 

The shadow can be characterized by using two observables (Hioki et al. 2009). The observable $R_s$ is defined as the radius of a reference circle passing by three points of the shadow: the top position $(\alpha_t, \beta_t)$, the bottom position $(\alpha _b,\beta _b)$, and the point corresponding to the unstable retrograde circular orbit seen by an observer on the equatorial plane $(\alpha _r,0)$. The distortion parameter $\delta _{s}$ is defined by the quotient $D/R_s$, where $D$ is the difference between the endpoints of the circle and of the shadow, both of them at the opposite side of the point $(\alpha _r,0)$. The radius $R_s$ gives the approximate size of the shadow, while $\delta_s$ measures its deformation with respect to the reference circle. If the inclination angle $\theta _{0}$ is independently known, measurements of $R_{s}$ and $\delta _{s}$ could serve to find the physical rotation parameter $A$ and the charge $Q$ (adimensionalized with $M$). These observables are given by $R_{s}=[(\alpha _t -\alpha_r)^2 + \beta_t ^2][2|\alpha _t -\alpha_r|]^{-1}$ and $\delta _s=(\tilde{\alpha}_p - \alpha_p)R_{s}^{-1}$, where $(\tilde{\alpha}_p, 0)$ and $(\alpha_p, 0)$ are the points where the reference circle and the contour of the shadow cut the horizontal axis at the opposite side of $(\alpha_r, 0)$, respectively.

In Fig. 1, the borders of the shadows of Kaluza-Klein black holes are shown for different values of the physical rotation parameter $A$ and the electric charge $Q$, with $0\le |Q| \le Q_{max}(A)$. The case $A=0$ is shown in the left plot; the size of the shadow decreases with $Q$, from $R_{s}=3\sqrt{3}$ until it shrinks to a point when $Q=Q_{max}(0)=2$. This is a remarkable feature of this theory, compared with the Reissner-Nordstr\"om General Relativity solution for which $R_{s}=3\sqrt{3}$ when $Q=0$, and reaching a minimum size $R_{s}=4$ in the extremal case $Q=1$. The center and right plots show the shadows corresponding respectively to $A=0.5$ and $A=0.9$; again, the size of them decreases with $Q$, starting from the same value as for the Kerr solution for fixed $A$ and $Q=0$, and reaching different extremal sizes for fixed $A$ and $Q=Q_{max}(A)$, when compared with those found for the Kerr-Newman solution. The maximum allowed charge for fixed $A$ is larger than for Kerr-Newman ones, so they can have larger amounts of charge before becoming naked singularities. The shadows of Kaluza-Klein black holes are always bigger than those of Kerr-Newman ones, for the same values of $A$ and $Q$. 
In Fig. 2 (left), the observable $R_{s}$ is plotted as a function of $Q$, for several values of $A$: it decreases with $Q$ for all $A$, and the values of $R_{s}$ are similar for the different values of $A$ considered in the plot; from the frame inside, where the range of $Q$ is smaller, it can be seen that $R_{s}$ increases with $A$. Each curve ends when the horizon fade out and a naked singularity is formed, for the value $Q_{max}(A)$. In Fig. 2 (center), the observable $\delta _s$ is plotted as a function of $Q$; it increases with the charge until a maximum distortion, obtained when $Q_{max}(A)$. The distortion is an increasing function of $A$ for a fixed value of $Q$. For the same values of $A$ and $Q$, the shadows corresponding to Kaluza-Klein black holes are less distorted than the shadows of Kerr-Newman ones. In Fig. 2 (right), the contour curves with constant $R_s$ and $\delta _s$ are shown in the plane $(A,Q)$; the gray zone representing naked singularities is outside the scope of this work, the boundary corresponds to the curve $Q_{max}(A)$. The values of $A$ and $Q$ an be extracted from the intersection of the curves with constant $R_s$ and $\delta _s$; there is no ambiguity because these curves intersect each other in a unique point. 
\begin{figure}[t!]
\begin{center}
\includegraphics[width=0.32\linewidth]{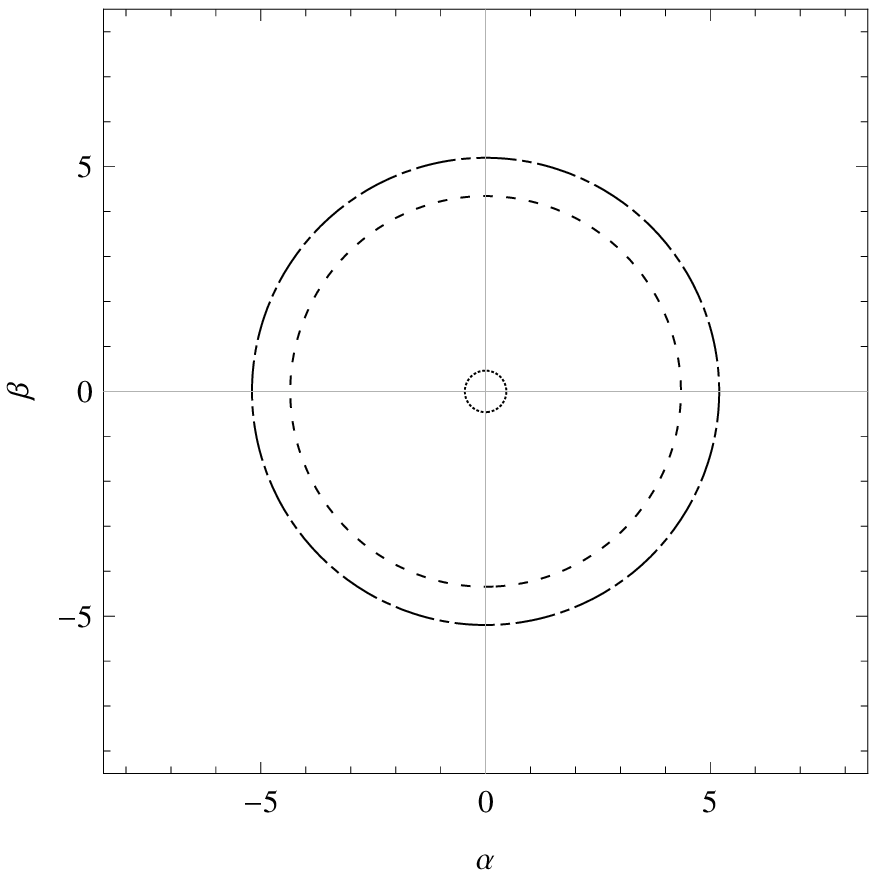}
\includegraphics[width=0.32\linewidth]{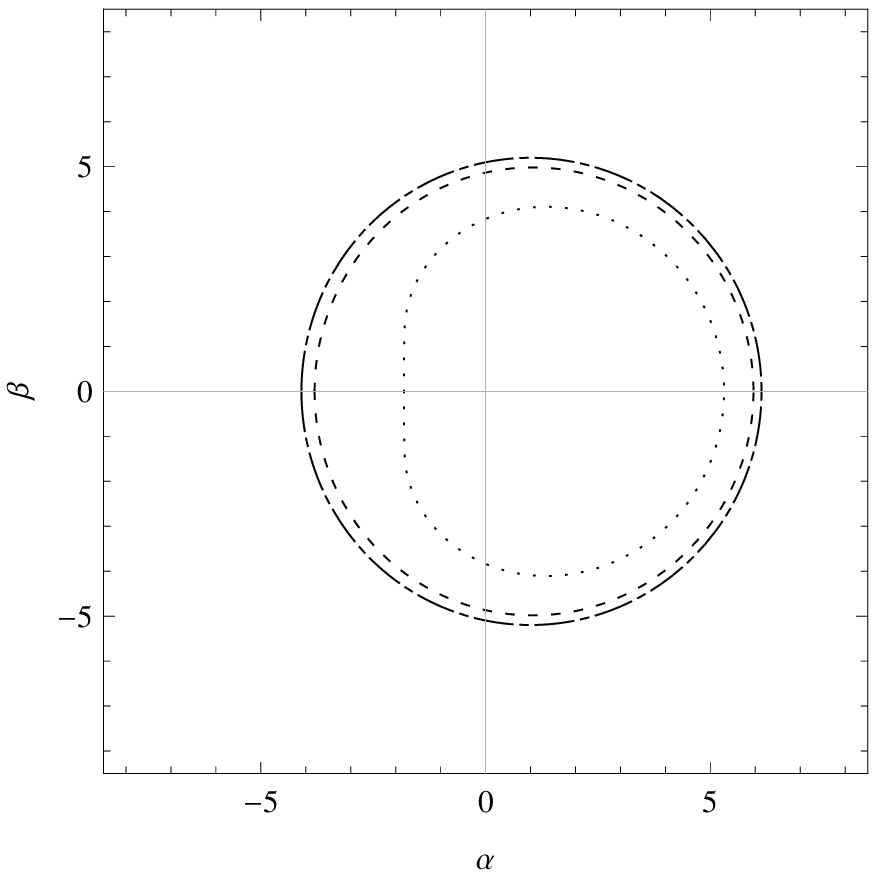}
\includegraphics[width=0.32\linewidth]{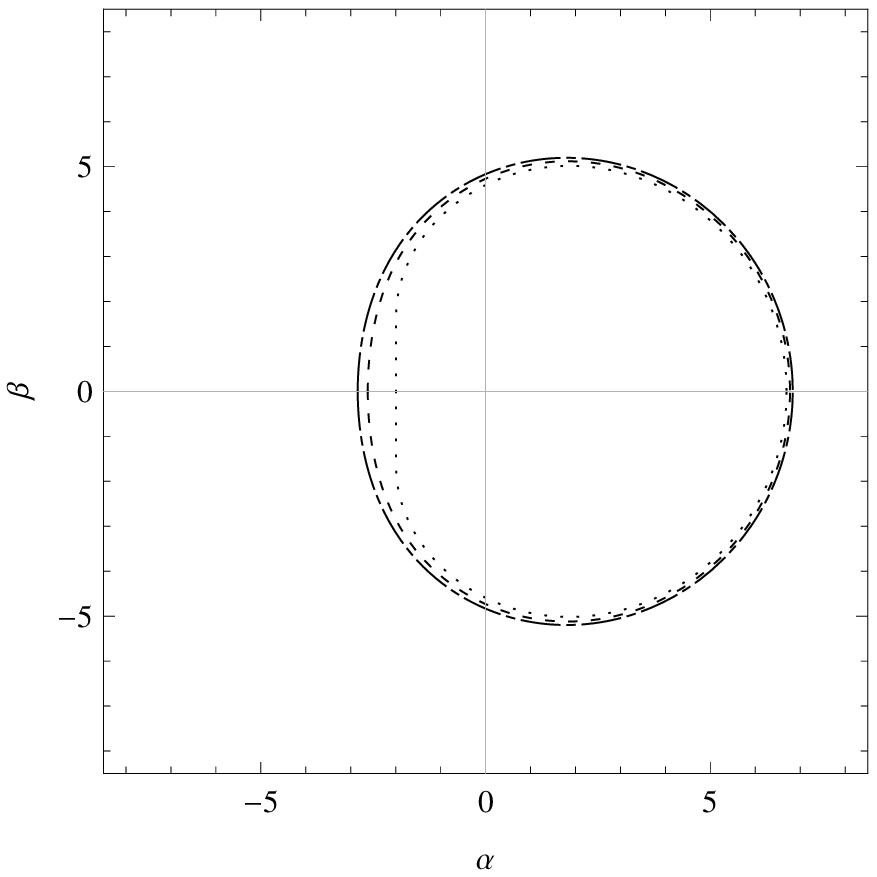}
\end{center}
\vspace{-0.5cm}
\caption{Shadow contours ($\theta _0=\pi /2$). Left: $A=0$, $Q=0$ (dashed-dotted), $0.5$ (dashed), and $1.99$ (dotted); center: $A=0.5$, $Q= 0$ (dashed-dotted), $0.5$ (dashed), and $Q_{max}=1.1298$ (dotted); right:  $A=0.9$, $Q=0$ (dashed-dotted), $0.3$ (dashed), and $Q_{max}=0.4583$ (dotted).}
\label{f1}
\end{figure}
\begin{figure}[t!]
\begin{center}
\includegraphics[width=0.31\linewidth]{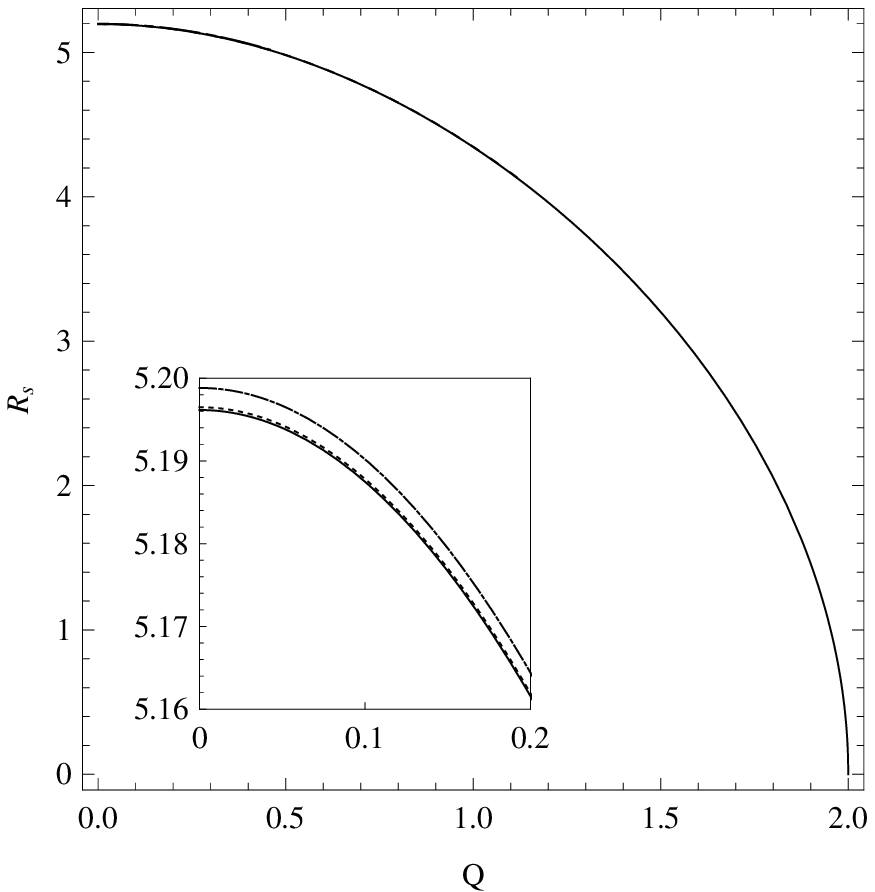}
\includegraphics[width=0.33\linewidth]{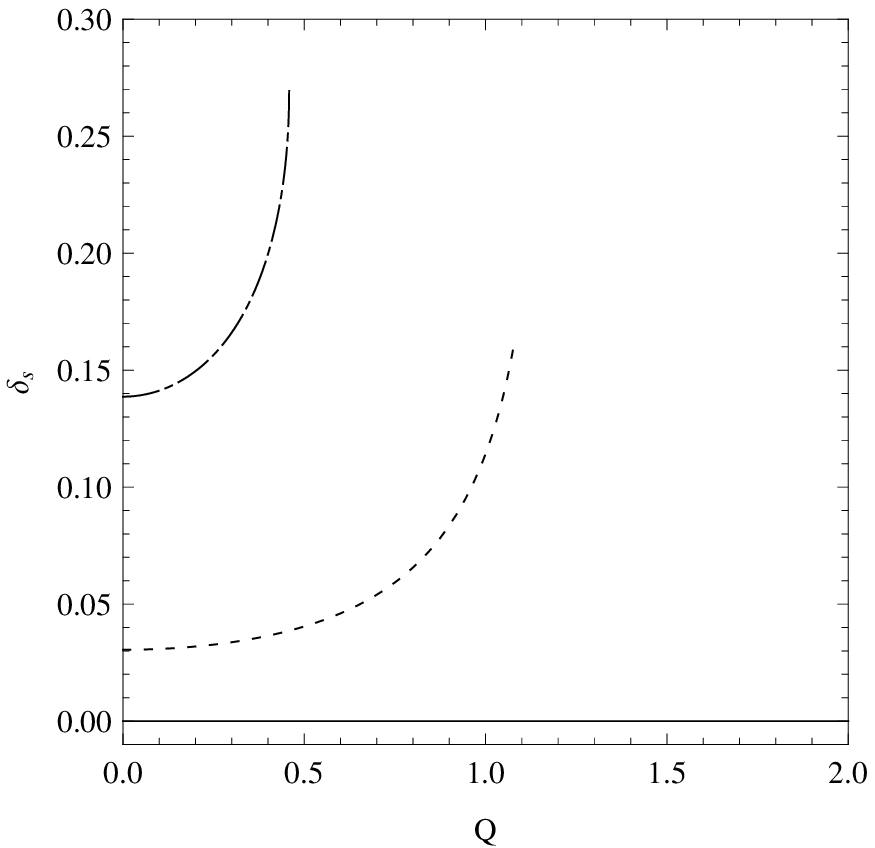}
\includegraphics[width=0.32\linewidth]{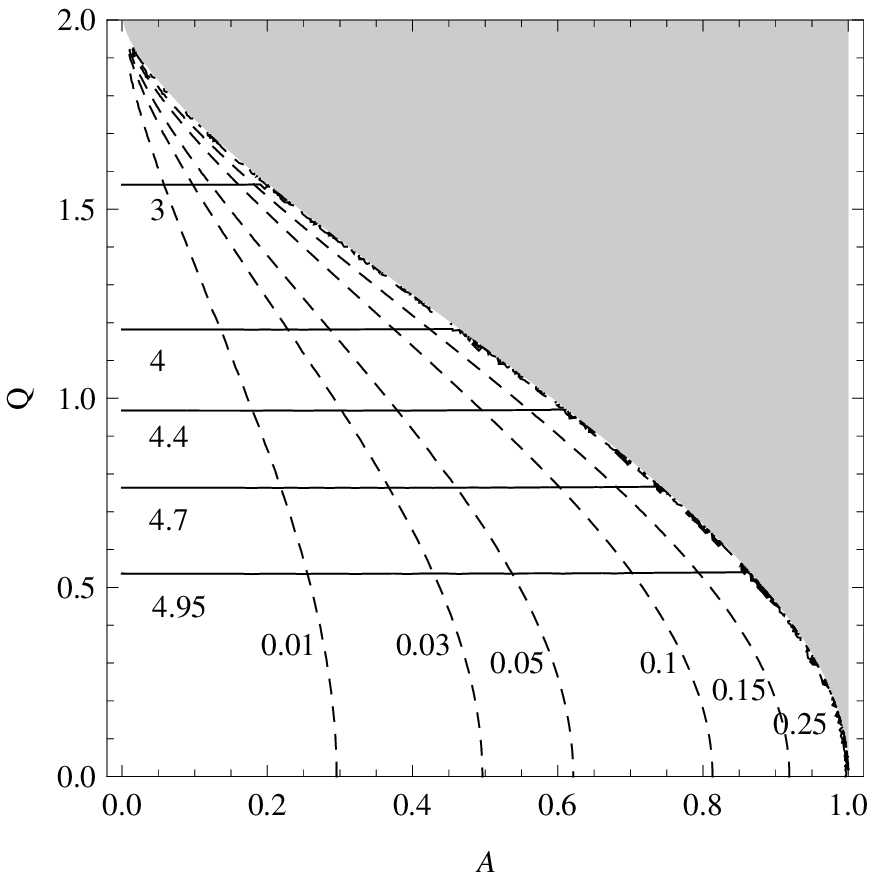}
\end{center}
\vspace{-0.5cm}
\caption{Observables ($\theta _0=\pi /2$). Left and center: plots of $R_{s}$ and $\delta _{s}$ for $A=0$ (full), $A=0.5$ (dashed), and $A=0.9$ (dashed-dotted). Right: contour plots of $R_{s}$ (full) and $\delta _{s}$ (dashed).}
\label{f2}
\end{figure}

The angular size of the shadow can be estimated by $\theta _s = R_{s}/D_{o}$, with $D_{o}$ the distance between the black hole and the observer. In the case of the supermassive Galactic black hole Sgr A* (Guillesen et al. 2009), for which $M=4.3 \times 10^{6}M_{\odot}$ and $D_{o}=8.3$ kpc, and taking $\theta_0=\pi/2$, the results are shown in the following table:
\begin{center}
\begin{tabular}{|c|c|c|c|c|c|c|}
\hline
& \multicolumn{3}{|c|}{$KN$}  & \multicolumn{3}{|c|}{$KKRD$}\\
\hline \hline
$ \;\;A=0 \;\;$ | $ \;\;Q \;\;$ & $0$ & $0.25$ & $0.5$ & $0$ & $0.25$ & $0.5$ \\
\hline
$\theta _s (\mu \mathrm{as})$ & $26.5718$ & $26.2916$ & $25.4047$ & $26.5718$ & $26.2959$ & $25.4763$ \\
\hline
$\delta _s (\%) $  & $0$ & $0$ & $0$ & $0$ & $0$ & $0$\\
\hline \hline
$ \;\;A=0.5 \;\;$ | $ \;\;Q \;\;$  & $0$ & $0.2$ & $0.4$ & $0$ & $0.2$ & $0.4$ \\
\hline
$\theta _s (\mu \mathrm{as})$ & $26.5735$ & $26.3951$ & $25.8419$ & $26.5735$ & $26.3968$ & $25.8707$ \\
\hline
$\delta _s (\%) $  & $3.05086$ & $3.19113$ & $3.69364$ & $3.05086$ & $3.18884$ & $3.64816$\\
\hline \hline
$ \;\;A=0.9 \;\;$ | $ \;\;Q \;\;$   & $0$ & $0.05$ & $0.1$ & $0$ & $0.05$ & $0.1$ \\
\hline
$\theta _s (\mu \mathrm{as})$ & $26.5855$ & $26.5744$ & $26.5413$ & $26.5855$ & $26.5745$ & $26.5414$ \\
\hline
$\delta _s (\%) $  & $13.8666$ & $13.9301$ & $14.1248$ & $13.8666$ & $13.9300$ & $14.1236$\\
\hline
\end{tabular}
\smallskip
\end{center} 
From the table one can see that a resolution of the order of $0.01$ $\mu \mathrm{as}$ or better is needed to observe deviations from General Relativity.

\section{Discussion}
\label{sec:discu}

We have investigated the shadow due to a spinning charged dilaton black hole, with coupling constant $\gamma =\sqrt{3}$, corresponding to a Kaluza-Klein reduction to four spacetime dimensions. We have obtained that, for fixed rotation parameter, mass, and charge, the presence of the dilaton results in a shadow that is slightly larger and with a reduced deformation, compared with the Kerr-Newman one.

In the next years, direct imaging of black holes will be possible (Johannsen et al. 2012). The Event Horizon Telescope, consisting of radio-telescopes scattered over the Earth, will reach a resolution of $15$ $\mu$as at $345$ GHz. RadioAstron is a space-based radio telescope launched in 2011, capable of carrying out measurements with $1$-$10$ $\mu$as angular resolution. The space-based Millimetron mission may provide the angular resolution of $0.3$ $\mu$as or less at $0.4$ mm. The MAXIM project is a space-based X-ray interferometer with an expected angular resolution of about $0.1$ $\mu$as. These instruments will be capable of observing the shadow of the supermassive Galactic black hole and those corresponding to nearby galaxies. However, in order to detect the deviations of General Relativity analyzed in this work, more advanced instruments with a better angular resolution is required.
 
\bigskip

\acknowledgments 
This work was supported by CONICET and UBA.


\begin{references}

\reference Amarilla, L. and Eiroa, E. F. 2013, Phys. Rev. D 87, 044057.

\reference Chandrasekhar, S. 1992, ``The mathematical theory of black holes'', Oxford U. P.

\reference Frolov, V. P., Zelnikov, A. I., and Bleyer, U. 1987, Ann. Phys. (Berlin) 499, 371.

\reference Guillesen et al. 2009, Astrophys. J. 692, 1075.

\reference Hioki, K. and Maeda, K.I. 2009, Phys. Rev. D 80, 024042.

\reference Horne, J. H. and Horowitz, G. T. 1992, Phys. Rev. D 46, 1340.

\reference Johannsen, T. et al. 2012, Astrophys. J. 758, 30.

\reference V\'azquez, S. E. and Esteban, E. P. 2004, Nuovo Cim. 119B, 489.


\end{references}
\end{document}